\documentclass[letterpaper]{aa}
\usepackage{txfonts}
\usepackage{natbib}
\usepackage{graphicx}
\bibpunct{(}{)}{;}{a}{}{,}

\begin{document}

\title{Detection of carbon monoxide in the high-resolution day-side spectrum of the exoplanet HD 189733b \thanks{Based on observations collected at the European Southern Observatory (186.C-0289)}}

\author{R.J.~de Kok\inst{\ref{int1}} \and M.~Brogi\inst{\ref{int2}} \and I.A.G.~Snellen\inst{\ref{int2}} \and J.~Birkby\inst{\ref{int2}} \and S.Albrecht\inst{\ref{int3}} \and E.J.W.~de Mooij\inst{\ref{int4}}}

\institute{SRON Netherlands Institute for Space Research, Sorbonnelaan 2, 3584 CA Utrecht, The Netherlands \email{R.J.de.Kok@sron.nl} \label{int1}  \and
Leiden Observatory, Leiden University, Postbus 9513, 2300 RA Leiden, The Netherlands \label{int2}  \and
Department of Physics, and Kavli Institute for Astrophysics and Space Research, Massachusetts Institute of Technology, Cambridge, Massachusetts 02139, USA \label{int3}  \and
Department of Astronomy and Astrophysics, University of Toronto, 50 St. George Street, Toronto, Ontario M5S 3H4, Canada \label{int4}  }

\date{}

\abstract
{After many attempts over more than a decade, high-resolution spectroscopy has recently delivered its first detections of molecular absorption in exoplanet atmospheres, both in transmission and thermal emission spectra. Targeting the combined signal from individual lines in molecular bands, these measurements use variations in the planet radial velocity to disentangle the planet signal from telluric and stellar contaminants.}{In this paper we apply high resolution spectroscopy to probe molecular absorption in the day-side spectrum of the bright transiting hot Jupiter HD 189733b.}{We observed HD 189733b with the CRIRES high-resolution near-infrared spectograph on the Very Large Telescope during three nights, targeting possible absorption from carbon monoxide, water vapour, methane and carbon dioxide, at 2.0 and 2.3 $\mu$m.}{We detect a 5-$\sigma$ absorption signal from CO at a contrast level of $\sim$4.5$\times$10$^{-4}$ with respect to the stellar continuum, revealing the planet orbital radial velocity at 154$^{+4}_{-3}$ km s$^{-1}$. This allows us to solve for the planet and stellar mass in a similar way as for stellar eclipsing binaries, resulting in $M_s =$ 0.846$^{+0.068}_{-0.049}$ $M_{\sun}$ and $M_p =$ 1.162$^{+0.058}_{-0.039}$ $M_{\mathrm{Jup}}$. No significant absorption is detected from H$_2$O, CO$_2$ or CH$_4$ and we determined upper limits on their line contrasts here.}{The detection of CO in the day-side spectrum of HD 189733b can be made consistent with the haze layer proposed to explain the optical to near-infrared transmission spectrum if the layer is optically thin at the normal incidence angles probed by our observations, or if the CO abundance is high enough for the CO absorption to originate from above the haze. Our non-detection of CO$_2$ at 2.0 $\mu$m is not inconsistent with the deep CO$_2$ absorption from low resolution NICMOS secondary eclipse data in the same wavelength range. If genuine, the absorption would be so strong that it blanks out any planet light completely in this wavelength range, leaving no high-resolution signal to be measured.}


\titlerunning{CO in the atmosphere of HD 189733b}
\authorrunning{de Kok et al.}

\maketitle 

\section{Introduction}

The extrasolar planet (exoplanet) transiting the K-star HD 189733 is one of the most studied exoplanets to date, since the brightness of the system (V=7.75 mag) allows the detailed study of its atmosphere through transit and secondary eclipse measurements. Optical transmission measurements show a steady decrease of the
effective planet radius from 0.3 to 1 $\mu$m, which is thought to be due
to Rayleigh scattering by a high altitude haze layer \citep{pon08,sin09,pon12}.  The Rayleigh slope seems only interrupted by an
absorption signal from sodium at 0.59 $\mu$m \citep{red08,hui12}. There is a strong ungoing debate on whether this
Rayleigh scattering signal continues in the near-infrared. Analysis of
HST NICMOS spectroscopic observations by \citep{swa08} show
absorption signatures from water and methane. However, reanalysis of the
same data \citep{gib11}, narrow band NICMOS observations \citep{sin09}, and HST data from WFC3 \citep{gib12} show no clear
evidence for molecular features and suggest that Rayleigh scattering
dominates the transmission spectrum up to 2.5 $\mu$m.

Numerous secondary eclipse measurements \citep[e.g.][]{dem06,gri08,cha08,swa09a} have placed constraints on the day-side temperature structure and trace gas abundances \citep[e.g.][]{mad09,lin12,lee12}. In
addition, phase curve observations from the Spitzer Space Telescope
suggest that the hottest point in the planet atmosphere is shifted with
respect to the sub-stellar point, indicating strong longitudinal
circulation in the atmosphere \citep{knu09,knu12}. However, also
several secondary eclipse measurements are controversial. \citet{swa10} claim the presence of an enormous emission feature at 3.25 $\mu$m
from ground-based observations, interpreted as non-LTE emission from
methane. The same team \citep{wal12} confirmed this result with
new data, in addition to showing similarly strong emission features at 2.3
$\mu$m. However, \citet{man11} see no sign of the L-band
emission in NIRSPEC data, with upper limits 40 times smaller than the
claimed non-LTE methane signal, and suggest that imperfect correction
for telluric water is the source of the feature claimed by Swain et al.
In any case, the unconventional observation and data analysis preformed
by Swain et al. and Waldmann et al. needs to be validated using other
techniques.

The current measurements of the day-side of HD 189733b are taken at low spectral resolution at best, with most observations being from broad-band photometry. The limited spectral resolution of the available measurements result in a large ambiguity in atmospheric composition \citep[e.g.][]{mad09,lin12,lee12} and temperature structure \citep{pon12}. 

Recently, high resolution spectroscopy has delivered its first
detections of molecular absorption: carbon monoxide was detected in the
transmission spectrum of HD209458b \citep{sne10} and in the
thermal dayside spectrum of $\tau$ Bo\"otis b \citep{bro12,rod12}. These observations, conducted with the CRyogenic
high-resolution InfraRed Echelle Spectrograph \citep[CRIRES][]{kau04} on the Very Large Telescope (VLT) at a resolution of
R$\approx$100,000 target the combined signal from individual lines in
the CO 2.3 $\mu$m molecular band. The large variation in the planet
velocity is used to disentangle its spectral features from the telluric and stellar contamination. This
also allows the determination of the orbital velocity of the planet, and
therefore the orbital inclination and true planet mass in the case of
non-transiting planets. Recent results from thermal dayside spectroscopy
of 51 Pegasi b \citep{bro13} are somewhat ambiguous. Although the first two nights of
data show a combined $\sim$6-$\sigma$ detection from carbon monoxide plus
water, no signal is detected in the third night. Although the signal on the third night was expected to be slightly weaker due to the reduced quality of the data and contamination by stellar lines, a complete absence was not expected based on the first two nights. This would mean that either some
yet unknown systematics is affecting the data, or that time variability (by e.g. global weather patterns), plays
a large role in the atmosphere of this object.

High-resolution spectroscopy has been attempted before for HD 189733 b. Using two nights of Keck/NIRSPEC observations at a resolution of $\sim$25,000, \citet{bar10} identify a possible planet signal using a deconvolution method. However, they could not make a strong claim due to the low significance of the result. Very recently, \citet{rod13} revisited this dataset and claim a $\sim$3-$\sigma$ detection of CO lines in HD 189733b.

Here, we present high-resolution CRIRES observarions of HD 189733 at 2.0 and 2.3 $\mu$m targeting absorption not only from CO, but also from CO$_2$, H$_2$O and CH$_4$ in the planet's day-side spectrum. Section 2 describes the observations and data analysis and Section 3 the cross-correlation procedure used to combine the signal from the individual lines. Section 4 describes the results, which are discussed in Section 5.

\section{Observations and data analysis}

We observed HD 189733b as part of a large programme \citep[ESO programme 186.C-0289,][]{sne11} to probe molecular signatures in hot Jupiter atmospheres with CRIRES. CRIRES consists of four 1024$\times$512-pixel Aladdin II detectors, separated by small gaps of roughly 100 pixels. The Multi-Application Curvature Adaptive Optics system \citep[MACAO,][]{ars03} was employed to maximise the throughput of the slit. Here, we use data taken during three nights, conducted at two wavelength settings, centred at 2.0 and 2.3 $\mu$m. The data were taken over a planet orbital phase-range near but not during secondary eclipse to maximise the illumination of the planet and the planet's change in radial velocity (see Table~\ref{table.obs} for more details). All observations used a 0.2" slit, resulting in a spectral resolution of R$\approx$100,000, and the star was nodded along the slit in an ABBA pattern to subtract the background.  The spectra around 2.0 $\mu$m contain possible planet lines of H$_2$O and CO$_2$, whereas the 2.3 $\mu$m region can be used to probe CO, H$_2$O and CH$_4$.

On 13 July data was taken with an integration time of 4$\times$15 seconds per nodding position. On 29 July the integration time was 5$\times$20 seconds and on 21 August 4$\times$20 seconds. All three observations spanned roughly 5 hours each.

   \begin{table}
      \caption[]{Details of the CRIRES observations, showing the observing date, the targeted walength range, the planet orbital phase, the number of spectra after background subtraction, and the net integration time.}
         \label{table.obs}
         \begin{tabular}{lcccc}
            \hline
            \noalign{\smallskip}
            Date      &  $\lambda$ ($\mu$m)  & Phase & $N$ & $t_{\mathrm{int}}$ (s) \\
            \noalign{\smallskip}
            \hline
            \noalign{\smallskip}
    	13 Jul 2011 & 2.2875-2.3452 & 0.3837-0.4801 & 110 & 13200 \\
	29 Jul 2011 & 1.9626-2.0045 & 0.5727-0.6661 & 69 & 13800 \\
	21 Aug 2011 & 1.9626-2.0045 & 0.3506-0.4377 & 76 & 12160 \\
            \noalign{\smallskip}
            \hline
         \end{tabular}

   \end{table}

The raw data was flat-fielded, background-subtracted, and optimally extracted using the standard CRIRES pipeline (version 1.11.0). The data from the three nights, for each of the four detectors, were analysed separately. In each case, a 1024$\times$$N$ matrix was constructed from the $N$ spectra. Bad columns in the matrix (corresponding to certain pixels in the spectra) were identified by taking the standard deviation of each column and finding $>$5-$\sigma$ outliers. These bad columns (of the order of 20 per detector) were masked in the rest of the analysis. Further individual bad pixels were replaced by spline-interpolated values.

The next step in the data reduction was to align all spectra onto a common wavelength grid. For this purpose we calculated a template telluric transmission spectrum with water, carbon dioxode and methane lines from HITRAN 2008 \citep{rot09} using line-by-line calculations with a single model layer and pressures, temperatures and gas concentrations relevant to Paranal. Since HD 189733A has significant CO lines in its spectrum, we also calculated CO absorption lines at the stellar temperature using HITEMP absorption data \citep{rot10} and scaled these relative to the telluric lines by comparisons with the data. We then Doppler-shifted the stellar CO lines according to the Earth's velocity, the known stellar systematic velocity and its planet-induced radial velocity variation \citep{tri09}, and combined them with the telluric spectrum. This was then correlated with each measured spectrum to find the best-fitting second-order polynomial correction on our initial guess of the wavelengths of each pixel, where the polynomial represents a wavelength-dependent shift. Finally, the spectra were spline-interpolated onto a common wavelength grid.

Before we searched for molecular features in the data using the cross-correlation technique, we first needed to remove telluric and stellar signatures. The telluric lines are static in position, but vary in strength due to changes in the geometric airmass and variations in the water vapour content of the atmosphere. The stellar lines are invariable in time to a high level, but slightly change in position during the night due to variations in the radial velocity component of the Paranal observatory with respect to the star (mainly caused by the Earth's rotation). 

\begin{figure}[htp]
\centering
\resizebox{0.8\hsize}{!}{\includegraphics{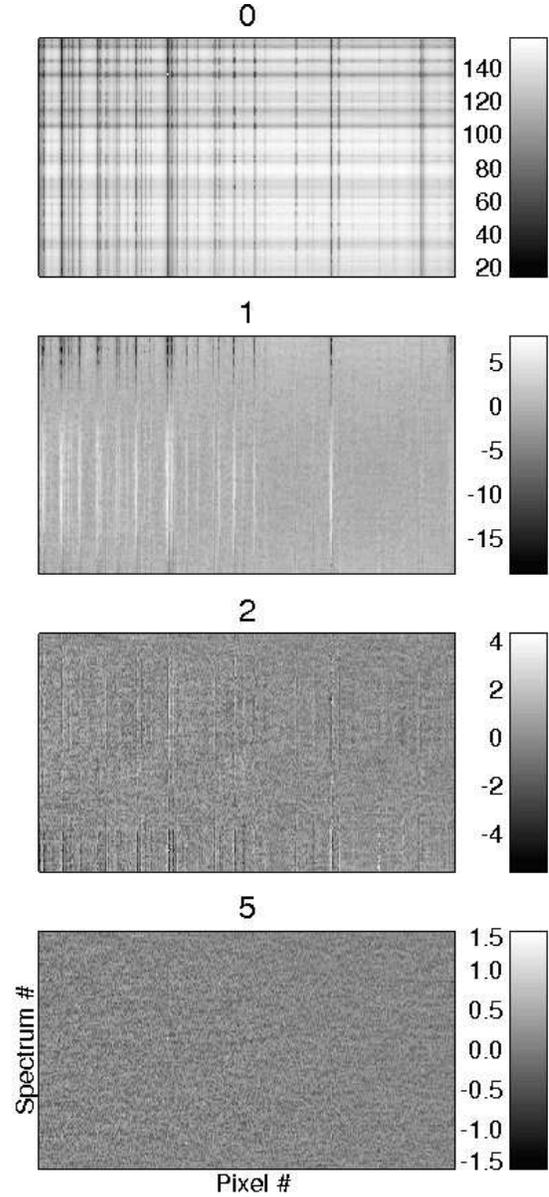}}
\caption{Example of stellar and telluric signal removal: the array of spectra (in ADU per second) for detector 2 at 2.3 $\mu$m on 13 July 2011 after setting the first $n$ singular values to zero, with $n$ being indicated above the panels.}
\label{fig.pcagrid}
\end{figure}

\begin{figure}[htp]
\centering
\resizebox{\hsize}{!}{\includegraphics{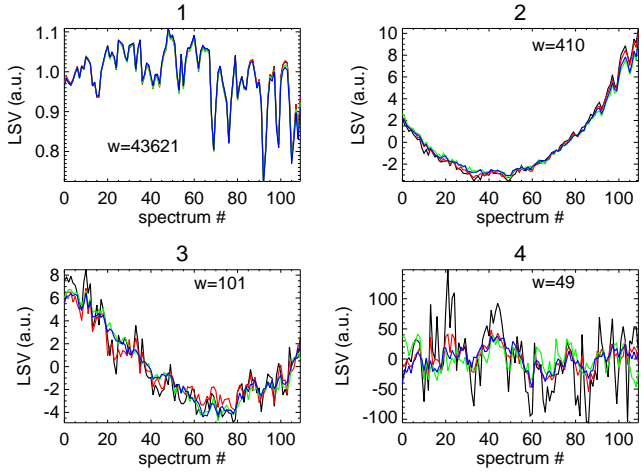}}
\caption{The first four left singular vectors for all four detectors (black -- detector 1; red -- detector 2; green -- detector 3; blue -- detector 4) at 2.3 $\mu$m on 13 July 2011, divided by their median value. The corresponding eigenvalues for detector 2 are also given to indicate their relative strengths. These vectors can be compared to the parameters in Fig.~\ref{fig.par}.}
\label{fig.eigu}
\end{figure}

\begin{figure}[htp]
\centering
\resizebox{\hsize}{!}{\includegraphics{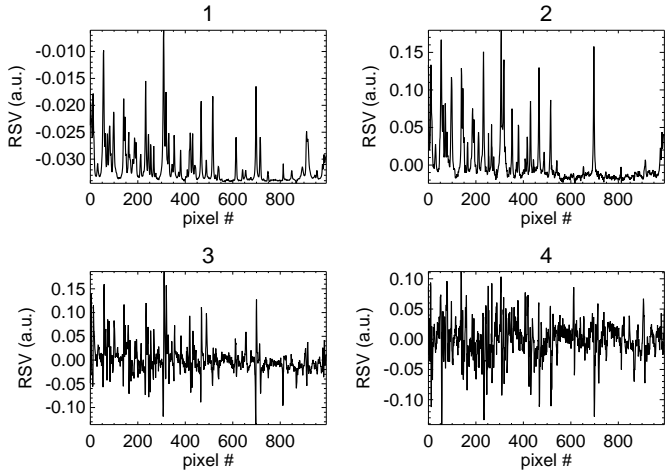}}
\caption{The first four right singular vectors, or eigenvectors, for detector 2 at 2.3 $\mu$m on 13 July 2011.}
\label{fig.eigenv}
\end{figure}

In this work we analyse the data in a somewhat different way than previously \citep{sne11,bro12,bro13} to explore new ways of reducing this type of data. We also performed the analysis with the previous method and found compatible results (see Section 4.1). Instead of removing the telluric contamination by fitting the columns in the matrix first with the geometric airmass and subsequently with the residual signals present in the strongest telluric lines, here we use singular value decompositions \citep[SVD, e.g.~][]{kal96}. For this we used the algorithm available in IDL. The SVD decomposes a matrix $A$ into the following matrices:

\begin{equation}
A=UWV^T,
\end{equation}

where $U$ is the matrix of left singular vectors, $W$ a (by definition) diagonal matrix of singular values and $V$ the matrix of right singular vectors. In our case the matrix $A$ represents the array of spectra on a single detector, with the number of pixels and number of spectra as dimensions. The right singular vectors are the eigenvectors and the left singular vectors, scaled by the singular values, indicate how each of the eigenvectors are weighted to form each observed spectrum. A convenient property of the SVD is that matrix $A$ can be approximated in an optimal way using only the largest few singular values, and setting the smaller singular values to zero \citep[e.g.][]{and75}. Here, we do the reverse and set the largest singular values to zero to remove the signals that are static with time while affecting the moving planet signal as little as possible. Fig.~\ref{fig.pcagrid} shows an example of a matrix of spectra (detector 2 on the night of 13 July 2011) with progressive subtraction of the singular values for detector 2 at 2.3 $\mu$m. Examples of the left and right singular vectors are shown in Figs.~\ref{fig.eigu} and \ref{fig.eigenv}.

\begin{figure}[htp]
\centering
\resizebox{\hsize}{!}{\includegraphics{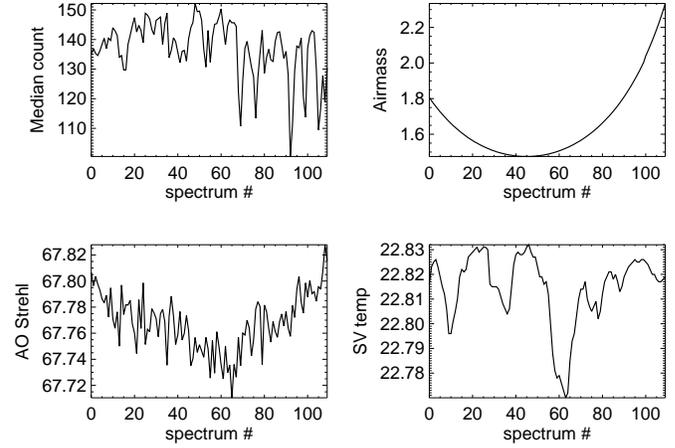}}
\caption{Instrument and environmental parameters possibly related to the left singular vectors plotted in Fig.~\ref{fig.eigu}. The median count is for detector 2 at 2.3 $\mu$m on 13 July 2011, AO Strehl indicates the Strehl ratio  from the adaptive optics system in percent and SV temp is one of the many temperature sensors on the instrument, located at the slit viewer.}
\label{fig.par}
\end{figure}

The left singular vectors corresponding to the first two singular values of the 2.3 $\mu$m data show clear correlations with the signal level in the continuum and the airmass for all four detectors (see Figs.~\ref{fig.eigu} and \ref{fig.par}). Hence, the subtraction of the first two singular values is in effect very similar to the first steps used by \citet{sne10} and \citet{bro12,bro13}. Also the third left singular vector is very similar for all four detectors. Its corresponding eigenvector (Fig.~\ref{fig.eigenv}) indicates it is related to shifts and/or broadening of the lines in the spectrum and its behaviour with time is somewhat similar to parameters relevant to the adaptive optics performance (see Fig.~\ref{fig.par}). The only parameter in the headers that was similar to the fourth left singular vector is one of the temperature sensors in the instrument. The physical correspondances of these third and fourth left singular vectors is far from conclusive, with correlation coefficients of around 0.5 between these vectors and the quantities plotted in Fig.~\ref{fig.par}, but it does highlight the diagnostic value of the SVDC for tracking potential systematic effects in the measurements. 
 
As a final step, we divide each column of the array of spectra by the square of the standard deviation. This prevents the cross-correlation analysis to be dominated by very noisy pixels.

\section{Correlation analysis}

We cross-correlated each reduced spectrum with model spectra of exoplanet atmospheres with Doppler shifts ranging from -200 to +200 km/s relative to the expected planet radial velocity. Orbital parameters of the system were taken from \citet{tri09}. In order to scan the parameter space and check for potential spurious correlation signals, we also performed our calculations for a wide range of amplitudes of the radial velocity of the planet, $K_p$ (50-200 km s$^{-1}$). We assumed a circular orbit throughout the paper, which is reasonable given that the maximum offset in planet radial velocity for its possible small eccentricity \citep[see][]{tri09} is smaller than the velocity resolution of CRIRES.

The model spectra are the result of line-by-line calculations using a range of parameterised pressure-temperature structures (see below), H$_2$-H$_2$ collision-induced absorption \citep{bor01,bor02} and absorption from a single trace gas. Line data for the trace gases are taken from HITEMP 2010 for H$_2$O and CO \citep{rot10}, HITRAN 2008 for CH$_4$ \citep{rot09} and HITEMP 1995 for CO$_2$ \citep{rot95}. A Voigt line-shape is used for calculation of the absorption. The atmospheric temperatures at pressures higher than 0.1 bar, where we take a temperature of 1350 K, are made to follow the best fitting temperature profiles of \citet{mad09} and the temperature at a lower pressure $p_1$ is varied from 500 to 1500 K in steps of 500 K. As such, we also included the option of weak thermal inversions at high altitudes. Between 0.1 and $p_1$ bar a constant lapse rate (i.e.~the rate of temperature change with log pressure) is assumed and $p_1$ is varied from 10$^{-1.5}$-10$^{-4}$ in multiplication steps of 10$^{0.5}$. At pressures lower than $p_1$ a uniform temperature is assumed. Volume mixing ratios of the gases ranged from 10$^{-6}$-10$^{-3}$ in multiplication steps of 10$^{0.5}$.

The correlation analysis was performed for each of the four detectors separately. To combine the correlation signals obtained from different spectra and detectors we first determined how to weigh their contributions. We did this by inserting an artificial planet signal in the wavelength-aligned data for the system velocity and $K_p$ under consideration, and perform the same data reduction as the original data. If the data with inserted signal gives a large increase in correlation value for a particular spectrum, this spectrum is thus expected to give a high signal also in the original data. Hence, the increase of the correlation when inserting the artificial planet signal can be used as a weighing factor to favour spectra or detectors for which a larger signal is expected. The correlation values for each spectrum and detector are multiplied by this increase of correlation, and added together to form the total correlation signal. 

In the correlation analysis for CO, there are still some residuals originating from the stellar spectrum. This is not surprising, since the star's radial velocity slightly changes with time due to the Earth's rotation and the star has significant CO lines in its spectrum. Fortunately, the velocities of the planet are at least 20 km/s offset from that of the star for the phases observed, so the strongest stellar residuals do not affect the correlation signal of the planet. When combining the correlation signals of the different spectra, we mask the velocities that are within 5 km/s of the stellar velocity after the cross-correlation. Note that this does not significantly affect our cross-correlation signals at the expected planet velocity, but only within the region of the expected stellar signals.

\begin{figure}[htp]
\centering
\resizebox{\hsize}{!}{\includegraphics{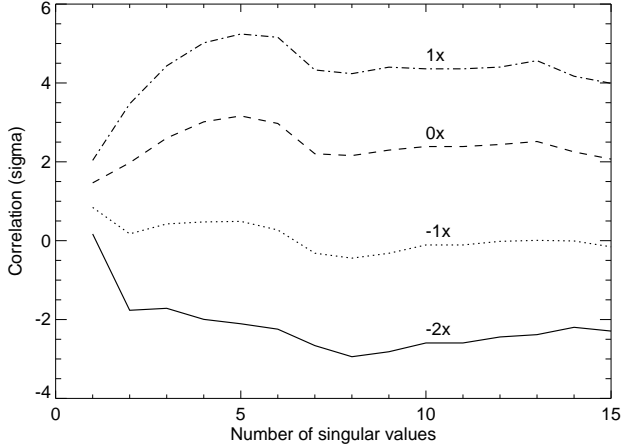}}
\caption{Total correlation at the planet position as a function of number of subtracted singular values for detector 2 at 2.3 $\mu$m on 13 July 2011 for the planet model giving the highest correlation. Different lines indicate different multiplication factors of the inserted planet signal (see text), as indicated by the labels.}
\label{fig.sigi}
\end{figure}

As more singular values are subtracted from the data, parts of the planet signal may also be removed. Hence, there is an optimum number of subtractions that will lead to the best recovery of the planet signal. We tested this effect by again inserting artificial planet signals and evaluating the correlation signal for the different numbers of reduced singular values. Fig.~\ref{fig.sigi} illustrates how the planet signal in a single detector first becomes apparent when the stellar and telluric signals are removed, and then slowly disappears as more and more singular values are removed. The slow drop of correlation with more subtracted singular values rises from the fact that there is only very little power in each of these higher singular values. The correlation as a function of number of subtracted singular values is actually very similar to the correlation of the data without inserted signal, but the correlation values are higher. For each detector, we chose the number of singular values to subtract according to the highest correlation of the inserted signal with respect to the correlation noise, when combined over the entire night. Fig.~\ref{fig.sigi} also illustrates what happens to the signal if negative planet signals are inserted. Such a negative signal would correspond to spectra with lines seen in emission instead of absorption.

\section{Results}

Our correlation analysis yields a detection of carbon monoxide at 2.3 $\mu$m, which is presented below. No significant signals are detcted for the other bands and molecules, of which the implications are discussed in Section 4.2.

\subsection{Detection of CO}

\begin{figure}[htp]
\centering
\resizebox{\hsize}{!}{\includegraphics{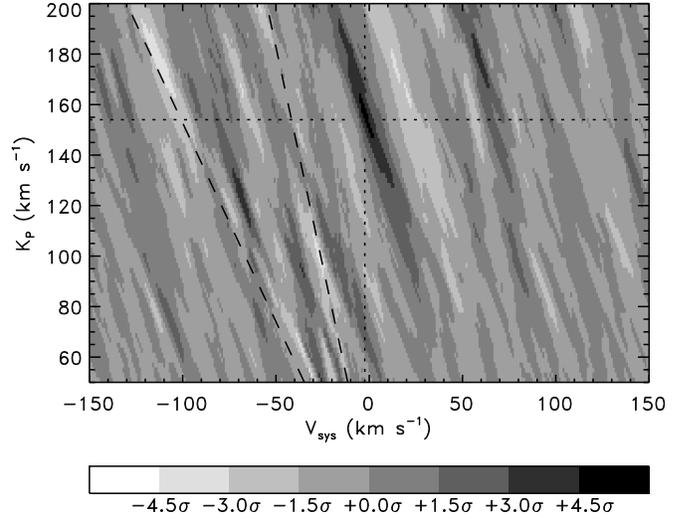}}
\caption{Combined cross-correlation values for all four detectors as a function of system velocity and $K_p$. The dashed line indicates where the stellar CO lines are expected to leave the largest contamination signal. The area in between the dashed lines can also be contaminated by stellar residuals. The dotted lines indicate the system velocity of HD 189733 and the peak position of $K_p$.}
\label{fig.combks}
\end{figure}

\begin{figure}[htp]
\centering
\resizebox{\hsize}{!}{\includegraphics{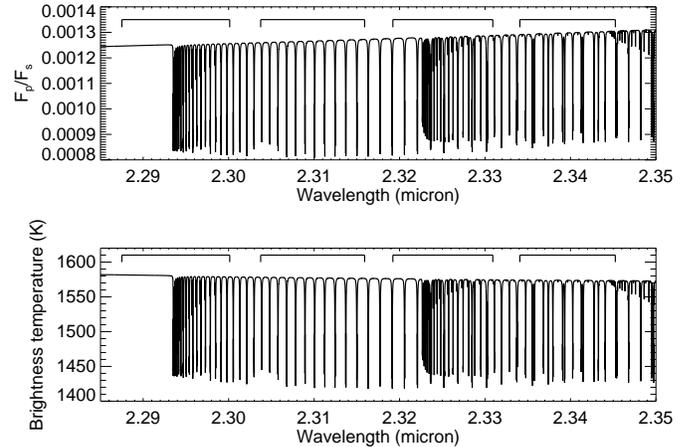}}
\caption{Template spectrum that gives the best correlation value. Note that we are not sensitive to the continuum level of the spectrum. For the sake of  clarity, stellar lines are ignored in the calculation of the planet-to-star contrast $F_p / F_s$. Brackets indicate the wavelength ranges covered by the four detectors.}
\label{fig.templ}
\end{figure}

The cross-correlation assuming different system velocities and values of the planet radial velocity $K_p$ for our best CO template are shown in Fig.~\ref{fig.combks}. The best CO template itself is shown in Fig.~\ref{fig.templ}. A positive correlation value (black) means in this case that the CO lines are apparent in absorption, as in Fig.~\ref{fig.templ}, indicating a non-inverted temperature profile over the probed pressure range. The correlation shows a 5.0-$\sigma$ maximum at the expected system velocity, within the range of $K_p$ expected from the measured radial velocity of the star and its estimated mass from spectral modelling \citep{tri09}. The significance was estimated by the peak cross-correlation value divided by the noise in the cross-correlation values in Fig.~6 at system velocities away from the planet signal. The signal is the combination of all four detectors, although detector 1 does not show a clear signal. Note however that detector 1 has a low weight, since artificially injected planet spectra give only very faint retrieved signals. This is consistent with the CO spectrum, which shows less lines in the wavelength range covered by detector 1. 

\begin{figure}[htp]
\centering
\resizebox{\hsize}{!}{\includegraphics{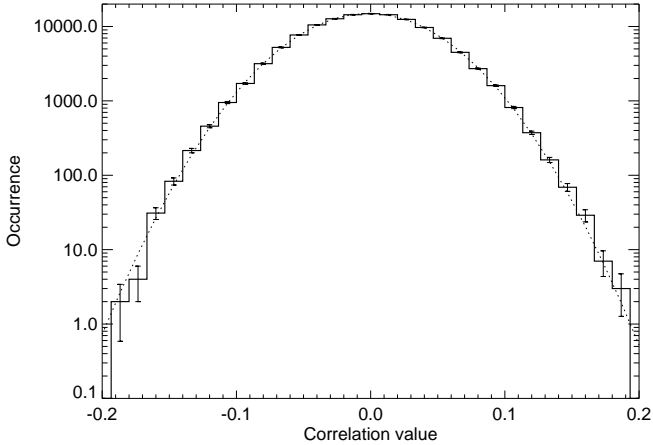}}
\caption{Histogram of correlation values outside the expected planet velocity (+10-400 km s$^{-1}$ with respect to the planet) for detectors 2-4 combined. Error bars are the suare root of the number of occurrences in each bin. The dotted line shows a Gaussian fit.}
\label{fig.hist}
\end{figure}

\begin{figure}[htp]
\centering
\resizebox{\hsize}{!}{\includegraphics{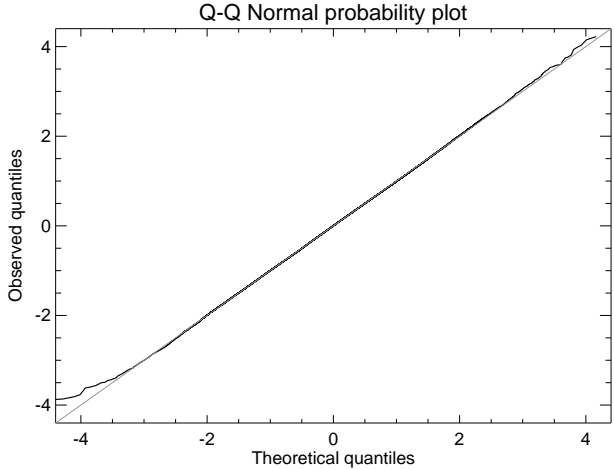}}
\caption{Quatiles of a normal distribution versus the measured quantiles of Fig.~\ref{fig.hist}. The gray line indicates a perfect normal distribution. The two agree well up till the 4-$\sigma$ level, which is determined by the number of available sampling points.}
\label{fig.qq}
\end{figure}

The observed CO signal is detected at $K_p = 154^{+4}_{-3}$ km s$^{-1}$ and $V_{\mathrm{sys}} = $ -2 km s$^{-1}$, both consistent with the literature values for the orbital and stellar parameters of the system \citep{tri09}. We estimated the statistical significance of the signal in several ways. First of all, we tested how the distribution of cross-correlation values as calculated for the 110 spectra compare to a Gaussian distribution. A histogram of the correlation values as obtained from detectors 2-4 is shown in Fig.~\ref{fig.hist}. The interval of the system velocities over shich the histrogram was plotted was chosen such that the area of the stellar signal was avoided and a large sample was obtainted. The best fit of the histogram to a Gaussian distribution is overplotted, showing a very good agreement. A further analysis of the quantiles of this distribution (Fig.~\ref{fig.qq}) shows it exhibits no significant deviation from Gaussianity. Only detector 1 shows some non-Gaussian behaviour using the optimal number of singular values, which improved when more singular values were subtracted. Since we did not find a significant signal in this detector and it is weighted very lightly compared to the other detectors, and because the extra singular values effectively only remove very little of the data, the additional subtraction of the singular values had no impact on the final result.

\begin{figure}[htp]
\centering
\resizebox{\hsize}{!}{\includegraphics{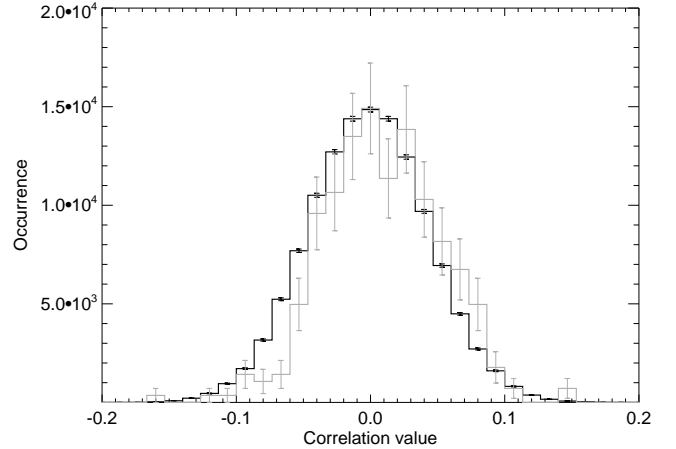}}
\caption{Histogram of correlation values away from the expected planet velocity (+10-400 km s$^{-1}$ with respect to the planet), and those at the planet velocity (gray line, scaled), for detectors 2-4 combined.}
\label{fig.hist2}
\end{figure}

An additional test we performed was the Welch T-test \citep{wel47}, comparing the planet signal distribution with the correlation distribution away from the expected planet velocities (see Fig.~\ref{fig.hist2}). Unfortunately, we have far less spectra available compared to \citet{bro12,bro13}, who performed the same test, so the sampling of the correlations at the planet velocity is worse here. Nevertheless, we can reject the null-hypothesis that the two samples are drawn from the same distribution at a 4.2-$\sigma$ confidence level.

Our measurements are not sensitive to the continuum level of the planet spectrum, since we remove any broadband signal in our data analysis. We are only sensitive to the depths of the narrow absorption  lines, which are the parts of the planet spectrum that show the Doppler shift. We determined the line contrast (the depth of the deepest lines with respect to the continuum, divided by the stellar flux) by inserting the model that gives the strongest signal times different (negative) scaling factors (see Section 3) until a significance value of 0$\pm$1 $\sigma$ was reached at the planet velocity for all four detectors combined. The line contrast for the deepest CO lines determined in this way is (4.5$\pm$0.9)$\times$10$^{-4}$, which corresponds roughly to the signal from the template spectrum (see Fig.~\ref{fig.templ}). This is the value for the model spectra and does not include the convolution to the instrument resolution. The line contrast including convolution gives a value of (3.4$\pm$0.7)$\times$10$^{-4}$. This line contrast can be interpreted as the difference of the line cores with respect to the broadband continuum and can be used as an additional constraint on the temperature profile and CO concentration (see the discussion in Section 5). The continuum level will be close to the broadband secondary eclipse depth around 2.3 $\mu$m for this transiting planet.

Using the data analysis method we presented previously \citep{sne10,bro12,bro13} we found a 4.8-$\sigma$ detection of the CO signal, which is consistent with the results presented here. Hence, either method can be used equally well, although the method presented here is perhaps less ad hoc and allows for a more objective treatment of the data.

\subsection{Upper limits for CO$_2$,H$_2$O and CH$_4$}

Besides CO, we also searched for signals from H$_2$O and CH$_4$ in the 2.3 $\mu$m data, and CO$_2$ and H$_2$O in the 2.0 $\mu$m data, but did not find a significant signal. We determined upper limits of the line contrast by inserting inverted models until a significance value of -3$\sigma$ was reached at the planet velocity for all four detectors combined. We used the same number of singular values as for the CO analysis. The resulting upper limits of the maximum model (unconvolved) line contrasts is 2.8$\times$10$^{-3}$ for H$_2$O and 8.3$\times$10$^{-4}$ for CH$_4$. The latter is comparable to the value of the medium-resolution secondary eclipse depth of 8$\times$10$^{-4}$, i.e.~the total planet-to-star flux contrast at these wavelengths, as measured with NICMOS on the HST by \citet{swa09a}. Since the continuum level would be close to the medium-resolution planet/star flux contrast, this would imply that the flux in the CH$_4$ lines would have to drop to zero for them to be detectable at the determined upper limit of the line depth. The contrast of the best-fitting CO lines is $\sim$ 4.5$\times$10$^{-4}$, so CH$_4$ would have to be significantly more abundant than CO for this to happen, given their similar line strengths in this region. Also, the upper atmosphere of HD 189733b would need to be extremely cold (of the order of a few hundred Kelvin). Neither of them is very likely \citep[see e.g.][]{mad09,mos13}. Hence, the upper limit on the line contrast of CH$_4$ does not constrain the abundance in a meaningful way if the measured medium-resolution planet-to-star contrast is correct. 

\begin{figure}[htp]
\centering
\resizebox{\hsize}{!}{\includegraphics{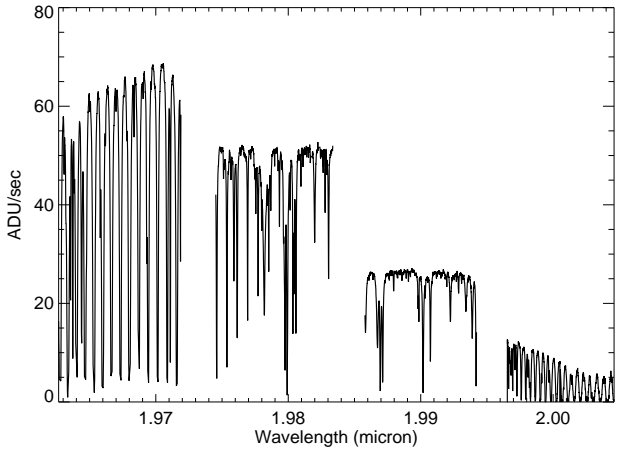}}
\caption{Example spectrum of the data at 2.0 $\mu$m.}
\label{fig.spec2}
\end{figure}

Although we have collected two nights of data at 2.0 $\mu$m, we were less sensitive to atmospheric signals in this wavelength range due to a combination of more telluric absorption and a smaller contrast between the planet and the star. Fig.~\ref{fig.spec2} shows an example spectrum of the 2.0 $\mu$m data. The spectra on detectors 1 and 4 are dominated by telluric CO$_2$ lines, greatly reducing the potential for retrieving exoplanet signals. In detectors 2 and 3 also water lines are evident. In general, the telluric transmission is significantly lower than around 2.3 $\mu$m. For this wavelength setting we ignored detector 4 due to the much lower count level. The SVD analysis could reduce the systematic signals down to similar levels as in Fig.~\ref{fig.pcagrid}. Again, models with varying temperature profiles and volume mixing ratios of H$_2$O and CO$_2$ were inserted with $K_p = 154$ km s$^{-1}$ to find 3-$\sigma$ upper limits of 8.7$\times$10$^{-4}$ for H$_2$O and 1.8$\times$10$^{-3}$ for CO$_2$. 

\subsection{Planet system parameters}

Since we now have a measure of the velocity and the orbital inclination of both the star and the planet around their common barycentre, model-independent estimates of their masses can be obtained in the same way as has been done for stellar eclipsing binaries for more than a century \citep[e.g.][]{tor10}. Using the values and uncertainties from \citet[][ see Table 2]{tri09} for the amplitude of the stellar radial velocity variations, planet orbital inclination and period, we determine a stellar mass of $M_s =$ 0.846$^{+0.068}_{-0.049}$ $M_{\sun}$ and a planet mass of $M_p =$ 1.162$^{+0.058}_{-0.039}$ $M_{\mathrm{Jup}}$. The stellar mass is consistent with that found by \citet{tri09} ($M_s =$ 0.823$^{+0.022}_{-0.029}$ $M_{\sun}$ and $M_p =$ 1.138$^{+0.022}_{-0.025}$ $M_{\mathrm{Jup}}$). In principle, if these CRIRES observations are repeated on both sides of the secondary eclipse (phases higher and lower than 0.5) and also at somewhat higher planet radial velocities (phases closer to 0.25 or 0.75), the uncertainty in stellar mass can be reduced to a level of $\sim$1\%. This issue is further discussed in Section 5.

   \begin{table}
      \caption[]{Stellar and orbital parameters taken from \citet{tri09} and derived from our measurements.}
         \label{table.pars}
         \begin{tabular}{lc}
            \hline
            \noalign{\smallskip}
            Parameter (unit) & Value  \\
            \noalign{\smallskip}
            \hline
            \noalign{\smallskip}
	    Assumed parameters (Triaud et al., 2009): & \\
$K_s$ (m s$^{-1}$) 	& 201.96 $^{+1.07}_{-0.63}$ \\
$P$ (days) 		& 2.21857312$^{+0.00000036}_{-0.00000076}$ \\
$i$ ($^\circ$)		& 85.508$^{+0.10}_{-0.05}$ \\
Measured parameters: & \\
$K_p$ (km s$^{-1}$)	& 154$^{+4}_{-3}$ \\
Derived parameters: & \\
$M_s$ ($M_{\sun}$) & 0.846$^{+0.068}_{-0.049}$ \\
$M_p$ ($M_{\mathrm{Jup}}$) & 1.162$^{+0.058}_{-0.039}$ \\
            \noalign{\smallskip}
            \hline
         \end{tabular}
   \end{table}

\section{Discussion and conclusions}

Using high resolution spectra from CRIRES on the VLT at orbital phases between 0.38-0.48, we detect carbon monoxide in the day-side atmosphere of HD 189733b at a 5-$\sigma$ significance. From Spitzer data the presence of CO was already suspected \citep{bar08,des09} , although more wider scans of the parameter space revealed no useful constraints on CO based on the Spitzer secondary eclipse data alone \citep{mad09}. HST/NICMOS secondary eclipse data do show large CO volume mixing ratios ($\sim$10$^{-4}$-10$^{-2}$) \citep{swa09a,mad09,lee12}, although past NICMOS results have been received with some scepticism \citep[e.g.][]{gib11}.  \citet{wal12} claim to detect a strong planet emission feature
at 2.3 $\mu$m using low-resolution groundbased spectroscopy, similar to
the controversial signal found at 3.2 $\mu$m by \citet{swa10}, and
also suggest it could be non-LTE emission. Due to the fact that
no proper physical model exists yet for this emission, we could not
perform a detailed search for this emission in our data. However, no
obvious high-resolution signal, expected from the $\sim$2x10$^{-3}$ low
resolution signal seen by Waldmann et al., is present in our data. Figure~1 indicates that single lines cannot be seen above the 1\% level in our data.

\begin{figure}[htp]
\centering
\resizebox{\hsize}{!}{\includegraphics{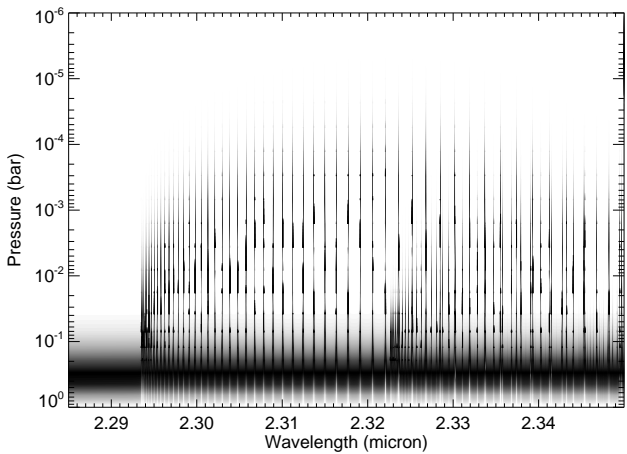}}
\caption{Normalised contribution functions for a model spectrum with CO at a volume mixing ratio of 10$^{-4}$. These show where in the atmosphere the thermal flux is emitted for each wavelength.}
\label{fig.cf}
\end{figure}

\begin{figure}[htp]
\centering
\resizebox{\hsize}{!}{\includegraphics{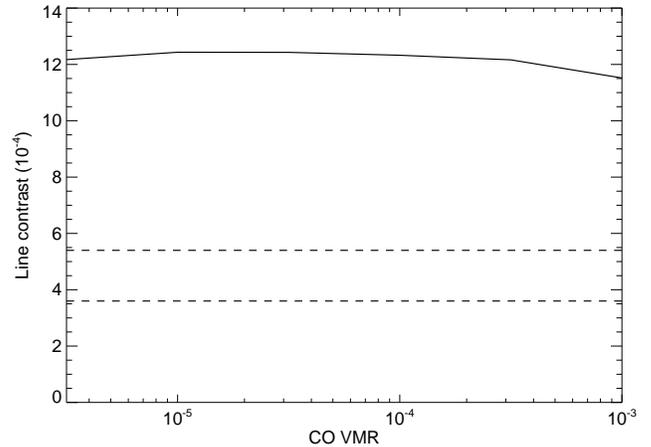}}
\caption{The maximum line contrasts (solid line) as a function of CO volume mixing ratio in our model grid (see Section 3). The minimum line contrast in a planetary atmosphere is zero for an isothermal atmosphere. Dashed lines show our determined limits on the CO line contrast.}
\label{fig.depths}
\end{figure}

The CRIRES measurements are not sensitive to the continuum level of the planet spectrum, since we remove any broadband signal in our data analysis. Instead, we determine an unconvolved line contrast (the depth of the deepest lines with respect to the continuum, divided by the stellar flux) of (4.5$\pm$0.9)$\times$10$^{-4}$ for the strongest CO lines. This line contrast indicates the difference between the thermal flux coming from the continuum level at high pressure and the flux from the line cores, which probe very low pressures (see Fig.~\ref{fig.cf}). The line contrast thus depends on the difference in temperatures at the pressures of the continuum and the line cores. Already it is clear that there is a large degeneracy in the temperature structure of the atmosphere that can reproduce a certain line contrast, since different sets of temperatures can have the same difference in flux. Furthermore, the abundance of CO adds another ambiguity, since increasing the CO abundance will decrease the pressure that is probed by the line cores. It is therefore uncertain at what pressure the flux of the line cores originates. This also means we cannot determine CO abundances from line contrasts alone. Fig.~\ref{fig.depths} illustrates this point. It shows the maximum line contrasts that were obtained in our model grid (see Section 3). Note that in this grid we kept temperatures at pressures above 0.1 bar fixed, meaning we basically fixed the continuum level, already reducing part of the possible degeneracies. Furthermore, our grid only covered a selected portion of the parameter space. A lower limit on the line contrast comes from an isothermal atmosphere, which would not show any lines (zero contrast). Fig.~\ref{fig.depths} shows that there is practically no dependence of the range of possible line contrasts on the CO volume mixing ratio. Hence, more information is needed before the CO abundance can be determined from these kinds of high spectral resolution measurements, most notably the temperature profiles covering pressures from the continuum level to the low pressures of the line cores. Constraints on the temperature profile can be obtained by measuring the thermal flux (or secondary eclipse depths) at a range of wavelengths, probing a range of pressures due to differences in gas absorption \citep[e.g.][]{mad09,lin12,lee12}. However, lower resolution measurements do not probe the low pressures of the line cores well. Also, there is still a degeneracy with absolute gas abundances, since this determines the pressure levels that are probed. Ideally, one would have a single gas for which we know (or can estimate reliably) the absolute abundance, and which probes a wide range of pressures from absorption lines or bands of different strengths. From measurements of this gas at different wavelengths the temperature structure can then be determined, if the variation of the gas abundance with pressure is known. CO$_2$ plays this role in the Venus atmosphere and CH$_4$ in the atmospheres of Jupiter, Saturn and Titan. Obtaining absolute gas abundances for giant exoplanet atmospheres is possible using transit (transmission) measurements if the H$_2$ Rayleigh scattering slope is measured \citep[e.g.][]{ben12}, although this signature might be difficult to distinguish from that of small haze particles. In summary, obtaining absolute gas abundances from line contrasts such as the ones we present here is difficult, but the line contrasts can be used as an additional constraint to reduce the uncertainty in retrievals. However, if multiple gases would be detected in a single high resolution observation, this would make it possible to determine the relative abundances of these gases, since they would both be reliant on the same temperature profile with the same continuum level at the same time. There might still be some degeneracy caused by the line cores of the two gases probing two different ranges of pressures, but these pressure ranges are likely to overlap.

In comparing our results with the claimed detection by \citet{rod13}, we note that the planet radial velocity measurements are consistent within their reported 1-$\sigma$ error bars. \citet{rod13} find a CO line contrast of 1.8$\times$10$^{-4}$ at a spectral resolution of R$=$25,000 with NIRSPEC at Keck. Assuming that the lines are unresolved, this translates to a line contrast of $\sim$7$\times$10$^{-4}$ at the CRIRES resolution, which is more than 50\% higher than what we find in our study. This may suggest that the reliability of the Keck signal is somewhat overestimated. However, the error bars on the line contrast of \citet{rod13} will probably be large, although these are not given in their paper. \citet{rod13} suggest a possible low CO abundance based on their analysis of the Keck data. Our line contrast is potentially even lower, but unfortunately we cannot yet draw conclusions on the CO abundance based on the line contrast alone (see the discussion above). Taking their 3-$\sigma$ detection using 13 hours of NIRSPEC observations at face value, the 5-$\sigma$ detection presented here implies that CRIRES is a factor 7 more efficient. We attribute this to the high spectral resolution and stability of CRIRES, and the well-performing adaptive optics system on the VLT.

Based on the available secondary eclipse data, retrieval methods favour temperature profiles that decrease in temperature as pressure decreases, i.e. a temperature profile without an inversion layer \citep{mad09,lin12,lee12}. However, in recent work \citet{pon12} argue that, based on their transmission measurements and improved analysis of secondary eclipse data, HD 189733b could also have temperatures that are increasing with decreasing pressure. This would then be the result of an optically thick haze layer, for which they see the signature in the transit measurements. However, our detection of CO shows that the temperature profile is not inverted at the pressures probed by the CO lines (10$^{-5}$-10$^{-3}$ bar for a CO volume mixing ratio of 10$^{-4}$, as seen in Fig.~\ref{fig.cf}, and roughly an order of magnitude lower pressures for a CO volume mixing ratio of 10$^{-3}$) and that the CO lines are not obscured by an optically thick haze layer. In principle it is possible that we are only seeing the cores of the CO lines sticking out above the haze in a higher part of the atmosphere where the temperatures decrease again with altitude, but this would require large CO volume mixing ratios ($> 10^{-4}$) if the temperature profile of \citet{pon12}, which shows an inversion down to at least 0.1 mbar, is correct. Such a high CO abundance with absorption features above the haze layer would then also be visible around 5 $\mu$m, where CO absorption is stronger than at 2.3 $\mu$m. Such a CO feature might indeed be visible in Spitzer data \citep{knu12}. An optically thick haze layer would also require a cold atmospheric layer above the haze layer, which is contrary to the conclusions of \citet{hui12}.  Hence, we conclude that the haze layer seen by \citet{pon08,sin09,pon12} is either optically thin when seen at normal incidence angles and that consequently there is no temperature inversion in the middle atmosphere, or that the haze layer is optically thick, but low enough for CO lines to be visible. 

We could derive only upper limits for the line contrasts of H$_2$O, CO$_2$ and CH$_4$: 8.7$\times$10$^{-4}$ for H$_2$O and 1.8$\times$10$^{-3}$ for CO$_2$ at 2.0 $\mu$m and 2.8$\times$10$^{-3}$ for H$_2$O and 8.3$\times$10$^{-4}$ for CH$_4$ at 2.3 $\mu$m. Future observations, also at more optimal wavelengths, will be needed to more effectively search for these gases and assess their relative abundances. Less telluric absorption and a higher planet-to-star contrast would be benificial in these observations. However, it is not straightforward to identify an optimal wavelength for these observations. For instance, the planet-to-star contrast is better at longer wavelengths, but there the sky background increases. Also, there is a trade-off between targeting strong lines in the planet spectrum, while avoiding strong lines in the telluric spectrum. A sensitivity study covering the CRIRES wavelength range, using realistic instrument performance and telluric absorption spectra, is needed to determine the optimal wavelengths for detection of these gases using high resolution spectroscopy. Note that the optimal observing wavelengths may differ for different targets.

From the measured radial velocity of $K_p = 154^{+4}_{-3}$ km s$^{-1}$ we derive a stellar mass of $M_s =$ 0.846$^{+0.068}_{-0.049}$ $M_{\sun}$ and a planet mass of $M_p =$ 1.162$^{+0.058}_{-0.039}$ $M_{\mathrm{Jup}}$, independent of stellar spectral modelling. These values are consistent with previously known values and their errors are somewhat larger. However, combining these results with possible future observations at phases larger than 0.5, as well as phases closer to 0.25 and 0.75, would help significantly in reducing the error on $K_p$, and hence in the masses of the star and planet \citep[e.g.][]{bro13}. 
An error in $K_p$ of 1 km s$^{-1}$ would result in an error on the stellar mass of 0.017 $M_{\sun}$, which is more precise than the current value derived from spectral modelling. The ability to treat transiting planet systems in the same manner as double-lined eclipsing binaries (DLEBs) allows a model-independent measurement of the stellar and planet masses and radii, which is particularly useful for planets with M-dwarf host stars. Precise dynamical measurements of the masses and radii of low-mass stars in DLEBs are not well-reproduced by stellar evolutions models and have radii that are up to $15\%$ larger than predicted \citep{lop05,rib06}. Strong magnetic fields in the short-period low-mass DLEBs are thought to be responsible for the radius inflation \citep[see e.g.][]{cha07,mor10}. Even in higher mass G- and K-dwarfs, strong magnetic activity in the stars results in disagreement with stellar models \citep{mor09}. However, in wider separation M-dwarf DLEBs with magnetically inactive low-mass stars, the radii are still inflated \citep{irw11,doy11}, with no real convergence towards stellar models at longer periods \citep{bir12}. Obtaining $K_p$ for planets transiting M-dwarfs removes the uncertainty in stellar models and gives a direct and accurate characterisation of the star-planet system.

\begin{acknowledgements}
This work was funded by the Netherlands Organisation for Scientific Research (NWO). We are thankful to the ESO staff of Paranal Observatory for their support during the observations and we thank the referee for his or her insightful comments.

\end{acknowledgements}

\end{document}